\documentclass[pre,onecolumn]{revtex4}
\usepackage{psfig}
\normalsize

\begin{document}

\title[Slow dynamics and local quasi-equilibrium]{SLOW DYNAMICS AND LOCAL QUASI-EQUILIBRIUM. RELAXATION IN SUPERCOOLED COLLOIDAL
SYSTEMS}

\author{J. M. Rub\'{\i}\footnote[3]{
Corresponding author: mrubi@ffn.ub.es}, I. Santamar\'{\i}a-Holek and A. P\'{e}rez-Madrid  
}

\affiliation{Department de F\'{\i}sica Fonamental, Facultat de F\'{\i}sica,
 Universitat de Barcelona, Diagonal 647, 08028 Barcelona, Spain}

\begin{abstract}
We present a Fokker-Planck description of supercooled colloidal systems exhibiting
slow relaxation dynamics. By assuming the existence of a local quasi-equilibrium
state during the relaxation of the system, we derive a non-Markovian Fokker-Planck
equation for the non-stationary conditional probability. \textbf{}A generalized
Stokes-Einstein relation \textbf{}containing \textbf{}the temperature of the
system at local quasi-equilibrium instead of the temperature of the bath \textbf{}is
obtained. \textbf{}Our results explain experiments showing that the
diffusion coefficient is not proportional to the inverse of the effective viscosity
at frequencies related to the diffusion time scale. \textbf{}
\end{abstract}
\maketitle

\section{Introduction}

The dynamics of slow relaxation systems exhibits peculiar characteristics which
make it essentially different from the dynamics of systems relaxing in much
shorter time scales \cite{angell}-\cite{nowak}. The aging behaviour of the
correlation functions \cite{struick6} and the violation of the fluctuation-dissipation
theorem \cite{kurchan} and of the Stokes-Einstein relation \cite{bonn6} are
among those significant features which have attracted the interest of many researchers
in the last few years. Relaxation phenomena in glasses, polymers, colloids and
granular matter provide innumerable situations demanding new theoretical \textbf{}developments
whose implementation constitutes \textbf{}a challenge for nonequilibrium statistical
mechanics theories. \textbf{}

The origin of that peculiar behaviour lies in the fact that during its evolution,
the system rests permanently away from equilibrium. This feature would justify
why results obtained when \textbf{}the system is close to equilibrium are not
necessary valid when that condition is not fulfilled. \textbf{}A simple example
illustrating this point is supplied from the slow dynamics of a two-level system.
During the transition, the system does not thermally equilibrate with the bath
and as a consequence equipartition law is not valid. It was shown in \cite{origin6}
that coarsening the level of description of the path that the system follows
in the configuration space going from diffusion to activation regimes leads
to violation of the fluctuation-dissipation theorem. The fast decay of the diffusion
modes makes the system always close to equilibrium whereas the activation process
occurring at longer time scales does not guarantee validity of the theorem.

In this paper we will present an example illustrating the peculiar dynamics
of slow relaxation systems. Using a Fokker-Planck description we derive a generalized
Stokes-Einstein relation for supercooled colloidal liquids showing that diffusion
coefficient and inverse of viscosity are not proportional to each other at low
frequencies, when the system becomes activated. At higher frequencies, the particles
undergo diffusion in the solvent and that relation holds. 

The paper will be distributed as follows. In Section \textbf{II}, we will briefly
review the generalization of the Onsager fluctuation theory to nonequilibrium
\emph{aging} states. Section \textbf{III} will be devoted to the formulation
of a mean field theory describing the dynamics of high concentrated suspensions.
This theory will be used in Section \textbf{IV} to interpret experimental results
supporting the violation of the Stokes-Einstein relation in supercooled colloidal
liquids. In the conclusions section we will summarize our main results.

\section{Slow relaxation dynamics and local quasi-equilibrium}

When the relaxation of a system is slow enough, its evolution takes place through
a sequence of states characterized by a set of thermodynamical quantities \textbf{}in
which the intensive parameters are, in general, different from those of the
bath. \textbf{}It is then said that the system is at quasi-equilibrium.

A statistical mechanics description of those systems involves the non-stationary
conditional probability density \( P(\underline{\alpha }_{0},t_{0}|\underline{\alpha },t) \),
with \( \underline{\alpha }=(\alpha _{1},...,\alpha _{n}) \) representing the
state vector of the system, and \( \underline{\alpha }_{0} \) the initial state
at time \( t_{0} \). The local quasi-equilibrium states are characterized by
the probability density \( P_{qe}(\underline{\alpha },t) \) satisfying

\begin{equation}
\label{limit P}
\lim _{t\rightarrow \infty }P(\underline{\alpha }_{0},t_{0}|\underline{\alpha },t)=P_{qe}(\underline{\alpha },t).
\end{equation}
 When relaxation takes place in a short time scale, \( P_{qe}(\underline{\alpha },t) \)
reduces to the local equilibrium probability density in which the intensive
parameters are those of the bath. 

The evolution of the conditional probability density is governed by a Fokker-Planck
equation. To obtain it, we will first formulate the continuity equation 
\begin{equation}
\label{continuidad alfa}
\frac{\partial }{\partial t}P(\underline{\alpha }_{0},t_{0}|\underline{\alpha },t)=-\frac{\partial }{\partial \underline{\alpha }}\cdot \left[ P(\underline{\alpha }_{0},t_{0}|\underline{\alpha },t)\underline{v}_{\underline{\alpha }}(\underline{\alpha }_{0},\underline{\alpha };t_{0},t)\right] ,
\end{equation}
in which \( P\underline{v}_{\underline{\alpha }} \) represents the phase space
probability current with \( \underline{v}_{\underline{\alpha }} \) the stream
velocity in \( \underline{\alpha } \)-space. That velocity can be obtained
from a mesoscopic version of the nonequilibrium thermodynamics for slow relaxation
systems we have proposed \cite{Slow6}.

The existence of a local quasi-equilibrium state in \( \underline{\alpha } \)-space
\textbf{}enables one to formulate the Gibbs equation 
\begin{equation}
\label{Gibbs ec}
T(t)\delta s_{qe}(t)=\delta e_{qe}(t)-\int \mu _{qe}(t)\delta P(t)d\underline{\alpha },
\end{equation}
 where \( s_{qe}(t) \) is the entropy and \( e_{qe}(t) \) the mean internal
energy per unit mass \textbf{}at quasi-equilibrium \textbf{}. We will assume
that the temperature \( T(t) \) \textbf{}of the system at local quasi-equilibrium
is only a function of time whereas the chemical potential \( \mu _{qe}(t) \)
can be a function of \( \underline{\alpha } \). \textbf{}

The entropy of the system can be expressed in terms of the probability density
by means of the Gibbs entropy postulate \cite{kampenSuper} in the form

\begin{equation}
\label{p. gibbs}
s(t)=-\frac{k}{m}\int P(t_{0}|t)\ln \frac{P(t_{0}|t)}{P_{qe}(t)}d\underline{\alpha }+s_{qe}(t),
\end{equation}
 where \( k \) is the Boltzmann constant and \( m \) the molecular mass. 

The form of the current \( P\underline{v}_{\underline{\alpha }} \) \textbf{}can
be inferred \textbf{}from the entropy production by relating it, as done in
nonequilibrium thermodynamics \cite{degroot6}, with the corresponding thermodynamic
force. The entropy production is obtained by combining the rate of change of
\( s_{qe}(t) \) obtained from Eq. (\ref{Gibbs ec}) with the time derivative
of Eq. (\ref{p. gibbs}). After using Eq. (\ref{continuidad alfa}) and integrating
by parts assuming that \textbf{\( P(t_{0}|t)\underline{v}_{\underline{\alpha }}(t_{0},t) \)}
vanishes at the boundary, one may identify the non-equilibrium chemical potential
\textbf{}\( \mu (t_{0};t)=\mu _{qe}(t)+\frac{kT(t)}{m} \ln \frac{P(t_{0}|t)}{P_{qe}(t)} \)
and the corresponding thermodynamic force \( \frac{\partial \mu (t)}{\partial \underline{\alpha }} \),
which is conjugated to the probability current \( P\underline{v}_{\underline{\alpha }} \).
Following the scheme of nonequilibrium thermodynamics one simply obtains

\begin{equation}
\label{stream vel}
\underline{v}_{\underline{\alpha }}(t_{0};t)=-\underline{\underline{B}}(t_{0};t)\cdot \left[ \underline{X}(\underline{\alpha },t)+\frac{kT(t)}{m}\frac{\partial }{\partial \underline{\alpha }}\ln P(t_{0}|t)\right] ,
\end{equation}
 where we have defined the generalized force \cite{degroot6}

\begin{equation}
\label{X force}
\underline{X}(\underline{\alpha },t)\equiv -\frac{\partial }{\partial \underline{\alpha }}\left[ \frac{kT(t)}{m}\ln P_{qe}(t)\right] .
\end{equation}

 The time dependent transport coefficients \( \underline{\underline{B}}(t_{0};t) \)
(related to the Onsager coefficients) incorporate memory effects through its
dependence on time \cite{Fp-nonmarkovian6}, \cite{oliveiraEURO}, and the local
quasi-equilibrium probability density is given by

\textbf{
\begin{equation}
\label{Pqe gral coll}
P_{qe}(t)=P_{e}e^{\frac{m}{k\, T(t)}[\mu _{qe}-e_{qe}]},
\end{equation}
} where \( P_{e} \) is the equilibrium distribution function. Eq. (\ref{Pqe gral coll})
can be obtained by using the statistical definition of the entropy in the Gibbs
equation (\ref{Gibbs ec}), \cite{Slow6}. Now, by substituting Eq. (\ref{stream vel})
into (\ref{continuidad alfa}) the resulting Fokker-Planck equation is 

\begin{equation}
\label{FP gral colloid}
\frac{\partial P(t_{0}|t)}{\partial t}=\frac{\partial }{\partial \underline{\alpha }}\cdot \underline{\underline{B}}(t_{0};t)\cdot \left[ \underline{X}(\underline{\alpha },t)P(t_{0}|t)+\frac{kT(t)}{m}\frac{\partial P(t_{0}|t)}{\partial \underline{\alpha }}\right] ,
\end{equation}

Using the Fokker-Planck equation (\ref{FP gral colloid}), we can calculate
the evolution equation for the equal-time correlation function of \( \underline{\alpha } \)-variables
which for sufficiently long times leads to the following expression for the
temperature of the system 

\begin{equation}
\label{T(t)}
T(t)\equiv \frac{m}{k}\left\langle \underline{X}(t)\cdot \underline{\alpha }(t)\right\rangle _{qe},
\end{equation}
with \( \left\langle ...\right\rangle _{qe} \) denoting \textbf{}average at
local quasi-equilibrium. This result reduces to the one corresponding to fast
relaxation processes \cite{degroot-pp88} in which the average is performed
at local equilibrium. 

In Ref. \cite{Slow6}, we have shown that the temperatures of the system and
of the bath \( T_{B}(t) \) are related by \( T(t)=AT_{B}(t) \), expressing
lack of thermal equilibration. Equilibrium is reached when \( A=1 \) in which
case, for linear thermodynamic forces, one recovers the expression for equipartition
law. The presence of \( T(t) \) in the last term of Eq. (\ref{FP gral colloid}),
gives rise to a modified version of the fluctuation-dissipation theorem.

\section{Fokker-Planck description of concentrated colloidal suspensions}

In this section, we will study diffusion in concentrated colloidal suspensions
by means of an effective medium approach elaborated on the grounds of the theory
we have introduced in the previous section. We will analyze the motion of a
test colloidal particle through the suspension by using the Fokker-Planck equation
describing the evolution of the two-time probability density \( P(\underline{\alpha }_{0},t_{0}|\underline{\alpha },t) \).
In this case, the phase space vector is simply \( \underline{\alpha }=(\vec{u},\vec{r}) \),
where \( \vec{u} \) and \( \vec{r} \) represent the velocity and position
of the \textbf{test} particle. 

To derive the Fokker-Planck equation, we will first formulate the Gibbs equation

\begin{equation}
\label{Gibbs colloids}
T\rho \delta s=\rho \delta e-\rho ^{-1}p\delta \rho -m\int \mu \delta Pd\vec{u},
\end{equation}
 which incorporates the effects of interactions among particles through the
term containing the excess of osmotic pressure \( p \). Here \( e \) is the
internal energy, \( m \) the mass of the particle and \( \rho =m\int Pd\vec{u} \)
the mass density. In this case, the quasi-equilibrium probability density is
given by \cite{Slow6}

\begin{equation}
\label{funcion equilibrio local}
P_{qe}=e^{\frac{m}{2kT}\left[ \mu _{qe}-\frac{1}{2}u^{2}-\rho ^{-1}p\right] }.
\end{equation}
 Here, the interaction of the test particle with the other colloidal particles
is represented by means of the term \( \rho ^{-1}p \) which can be expressed
in terms of a virial expansion in \( \rho  \), whereas \textbf{}\( \mu _{qe} \)
constitutes the ideal chemical potential. Introducing the fugacity \( z=P\, a \),
with the activity coefficient given by \( a\equiv e^{\frac{m}{kT}\rho ^{-1}p} \),
Eq. (\ref{p. gibbs}) can be expressed as

\begin{equation}
\label{post Gibbs fugacity}
s=-\frac{k_{B}}{\rho }\int P\, \ln \frac{z}{z_{qe}}d\vec{u}+s_{qe},
\end{equation}
 where \( z_{qe}=e^{\frac{m}{kT}[\mu _{qe}-\frac{1}{2}u^{2}]} \) is the fugacity
at quasi-equilibrium. Notice that in the limit of infinite dilution \( \rho ^{-1}p\cong 0 \),
thus implying \( a=1 \), which corresponds to the ideal case. \textbf{}

The existence of interactions among the test and the other colloidal particles,
gives rise to a force term in the continuity equation for \( P \). In particular,
according to Eq. (\ref{funcion equilibrio local}), the force entering into
this term is given by  \textbf{\( -\nabla (\rho ^{-1}p) \)} (see for example
Ref. \cite{mayorgas}). Now, by taking into account the definition of the activity
coefficient \( a \), the continuity equation for \( P \) becomes \textbf{}

\begin{equation}
\label{continuidadch6}
\frac{\partial P}{\partial t}+\nabla \cdot P\vec{u}-\nabla \left[ \frac{kT(t)}{m}\ln a\right] \cdot \frac{\partial P}{\partial \vec{u}}=-\frac{\partial }{\partial \vec{u}}\cdot P\vec{v}_{\vec{u}},
\end{equation}
 where \( P\vec{v}_{\vec{u}} \) is the probability current, with \( \vec{v}_{\vec{u}} \)
the stream velocity in \( \vec{u} \)-space. 

In accordance with the general formalism of nonequilibrium thermodynamics, the
expression of the probability current \( P\vec{v}_{\vec{u}} \) can be obtained
from the entropy production which only contains dissipative terms and is simply
given by

\begin{equation}
\label{sigma colloid}
\sigma =-\frac{m}{T}\int P\vec{v}_{\vec{u}}\cdot \frac{\partial \mu }{\partial \vec{u}}d\vec{u}.
\end{equation}
Eq. (\ref{sigma colloid}) has been obtained by taking the time derivative of
Eq. (\ref{post Gibbs fugacity}), using Eqs. (\ref{continuidadch6}), (\ref{funcion equilibrio local})
and (\ref{Gibbs colloids}), and the balance equation for the energy \( e(t) \),
neglecting viscous dissipation \cite{degroot6}. In order to obtain Eq. (\ref{sigma colloid}),
we have also identified the nonequilibrium chemical potential 

\begin{equation}
\label{mu colloids 6}
\mu (t_{0};t)=\mu _{qe}(t)+\frac{kT(t)}{m} \ln \frac{z(t_{0};t)}{z_{qe}(t)}.
\end{equation}

By taking into account the definition (\ref{X force}), the resulting generalized
Fokker-Planck equation for the two-time probability density is 

\begin{equation}
\label{FP-compres}
\frac{\partial P}{\partial t}+\nabla \cdot \vec{u}P-\nabla \left[ \frac{kT(t)}{m}\ln a\right] \cdot \frac{\partial P}{\partial \vec{u}}=\frac{\partial }{\partial \vec{u}}\cdot \beta (t)\left[ \vec{u}P+\frac{kT}{m}\frac{\partial P}{\partial \vec{u}}\right] ,
\end{equation}
 where \( \beta  \) characterizes the dissipation of the kinetic energy of
the test particle. This coefficient is, in general, function of time, position
and \textbf{}density; \textbf{}its dependence on time \textbf{}is a consequence
of the non-Markovian nature of the stochastic process \cite{Fp-nonmarkovian6},
\textbf{}\cite{oliveiraEURO}, \cite{tokuyama}. \textbf{}

From Eq. (\ref{FP-compres}), we will now derive the evolution equation for
the velocity field by taking the time derivative of the momentum \( \rho \vec{v}=m\int \vec{u}Pd\vec{u} \)
and using Eq. (\ref{FP-compres}). After integrating by parts one obtains

\begin{equation}
\label{momentum supercool}
\rho \frac{d\vec{v}}{dt}=-\beta \rho \vec{v}-\frac{kT(t)}{m}\left[ 1+\frac{\partial \ln a}{\partial \ln \rho }\right] \nabla \rho ,
\end{equation}
 where we have used the fact that, in the absence of an externally imposed flow,
the second moment \( \! \! \, \, \vec{\vec{\mathrm{P}}}\! \! \, \, =m\int (\vec{u}-\vec{v})(\vec{u}-\vec{v})Pd\vec{u} \)
can be approximated by \( \! \! \, \, \vec{\vec{\mathrm{P}}}\! \! \, \, =\frac{kT}{m}\rho \vec{\vec{1}} \),
\cite{nosotrosPRE6}.

\begin{figure}
\begin{center}
\mbox{\psfig{file=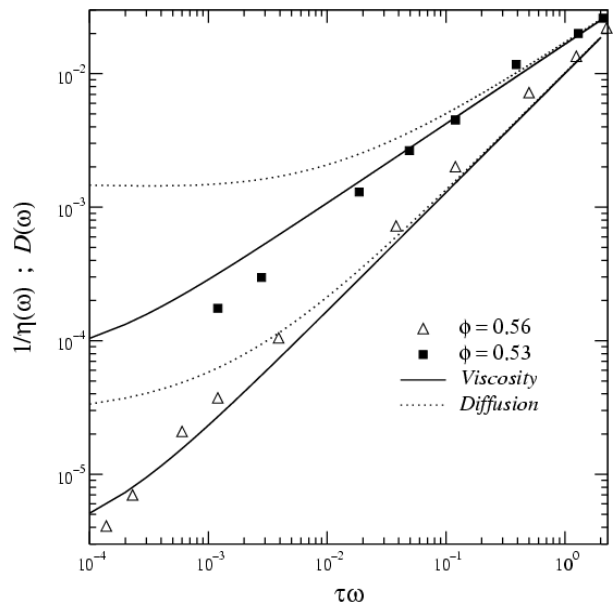,width=8cm}}
\end{center}
\caption{\label{ViscDiffCH6A}\small Comparison between the viscosity and diffusion coefficients for two values 
of the volume fraction. Triangles and squares represent viscosity data from Ref.
\cite{bonn6}. The gap between viscosity and diffusion at short frequencies is proportional to the 
scaling factor $A$ of the temperature. At larger frequencies $A \rightarrow 1$.
}
\end{figure}

At times such that \( t\gg \beta ^{-1} \), the particle enters the diffusion
regime in which one may neglect the time derivative in Eq. (\ref{momentum supercool}),
obtaining then the constitutive relation for the mass diffusion current \( \rho \vec{v} \)
which substituted into the mass continuity equation \( \frac{\partial \rho }{\partial t}=-\nabla \cdot (\rho \vec{v}) \),
yields the Smoluchowski equation 

\begin{equation}
\label{Smoluchowski}
\frac{\partial \rho }{\partial t}=\nabla \cdot \left( D\cdot \nabla \rho \right) .
\end{equation}
From Eq. (\ref{momentum supercool}) we may identify the diffusion coefficient

\begin{equation}
\label{Diff coef}
D=\frac{kT(t)}{m\beta }\left[ 1+\frac{\partial \ln a}{\partial \ln \rho }\right] .
\end{equation}
 The form of Eq. (\ref{Diff coef}) coincides with the expression given in Ref.
\cite{degroot6}. Notice, however, that it contains the temperature of the system
at quasi-equilibrium instead of the temperature of the bath.

\subsection{The generalized Stokes-Einstein relation in supercooled colloidal systems}

It has been shown in \cite{bonn6} that in supercooled colloidal liquids at
short frequencies the Stokes-Einstein relation is not fulfilled. Measurements
of the viscosity and the diffusion coefficients obtained in Refs. \cite{mortensen6},
\cite{mason6}, indicate the existence of a frequency domain in which those
quantities are not proportional (see figure 1). It is our purpose to offer here
an explanation for those experimental results. 

In \cite{mortensen6} it was shown that the colloidal system can be \textbf{}assumed
isotropic and homogeneous at the considered volume fractions, ranging from the
freezing to the glass transition values \textbf{\( \phi _{g} \)}. Consequently,
the transport coefficients can be assumed to be independent of the wave vector
\( \vec{k} \). We will first introduce \textbf{}the \textbf{}effective viscosity \textbf{}

\begin{equation}
\label{D(w) colloid}
\eta (\omega )=\frac{kT}{6\pi R}D^{-1},
\end{equation}
 \textbf{}where \( R \) is the radius of the particle. 

The dependence of the viscosity on the frequency can be inferred through the
expression of the penetration length \( \lambda  \) of the diffusive \textbf{}modes.
\textbf{}In the dilute case, \( \lambda  \) is given through \( \lambda ^{-1}=\frac{1}{R}\left( \tau \omega \right) ^{\frac{1}{2}} \),
with \( \tau =\frac{R^{2}}{6D_{0}} \) the characteristic diffusion time \cite{boon-yip}
and \textbf{\( D_{0}=\frac{kT_{B}}{6\pi R\eta _{0}} \)} the diffusivity in
the pure solvent, \textbf{}with \( \eta _{0} \) its viscosity. At higher concentrations,
it can be assumed to be

\begin{equation}
\label{lambda concentrated}
\lambda ^{-1}\equiv \frac{1}{\Gamma R}\left( \tau \omega \right) ^{\gamma },
\end{equation}
 where the exponent \( \gamma  \) and the scaling function \( \Gamma  \) may,
in general, depend on the volume fraction. The quantity \( \Gamma R \) can
be interpreted as an effective radius. \textbf{}

The penetration length gives an estimation for the average size of a cage formed
by the surrounding particles \cite{cohen6}. At high frequencies, the test particle
performs a free Brownian motion inside the cage and the viscosity is that of
the pure solvent. At short frequencies, collisions of the test particle with
the other particles modify the viscosity. \textbf{}Thus, we will assume the
following expression for the effective viscosity \textbf{}

\begin{equation}
\label{eta w lambda}
\eta (\omega )=\eta _{0}\left( 1+R\lambda ^{-1}\right) ,
\end{equation}
valid up to first order in \( R\lambda ^{-1} \) reflecting the fact that cage
diffusion is important when \( R\sim \lambda  \), \cite{cohen-cage}. \textbf{}

By taking into account the general relation between system and bath temperatures
obtained from our formalism, \( T=AT_{B} \), \cite{Slow6}, \textbf{}Eq. (\ref{D(w) colloid})
can be written as \textbf{}

\begin{equation}
\label{SE super}
D\eta =\frac{kT_{B}}{6\pi R}A,
\end{equation}
 which constitutes a generalization of the Stokes-Einstein relation to systems
exhibiting slow relaxation dynamics. 

From Eq. (\ref{T(t)}) and from its definition, the quantity \( A \) is essentially
proportional to the velocity moments of the particle. In particular, when \( \underline{X}(t) \)
is a linear function, \( A \) is proportional to the velocity correlation function.
It is then plausible to assume the following expansion \cite{ngai}, \cite{tokuyamaslow} \textbf{}

\begin{equation}
\label{A(w)}
A\cong \left[ 1+b\left( \tau \omega \right) ^{\epsilon }\right] ,
\end{equation}
 where \( b \) and the exponent \( \epsilon  \) may, in general, be functions
of the volume fraction \( \phi  \). Comparison of Eq. (\ref{A(w)}) with the
corresponding relation obtained in \cite{bonn6} by fitting the experimental
data yields \( \ln b\cong \epsilon \left( 40\frac{\phi }{\phi _{g}}-37\right) +2.659 \),
and \( \epsilon =-0.77 \). 

To contrast the theoretical expression of the effective viscosity (\ref{eta w lambda})
\textbf{}with the experimental results reported in \cite{bonn6}, we will take
the logarithm of Eq. (\ref{eta w lambda}) and expand the result in terms of
the variable \( \log \tau \omega  \), around \( \tau \omega =1 \). Up to first
order we obtain

\begin{equation}
\label{log eta short}
\log \eta \cong \log \left[ \eta _{0}\left( 1+\frac{1}{\Gamma }\right) \right] +\frac{\gamma }{1+\Gamma }\log \tau \omega .
\end{equation}
 A direct comparison of Eq. (\ref{log eta short}) with the fitting of experimental
data given by the expression \( \log \eta \cong c_{0}+s(\phi )\log \tau \omega  \),
with \( s(\phi )\approx 6.39-13\phi  \) and \( c_{0} \) a constant, gives
the dependence of \( \gamma  \) and \( \Gamma  \) on the volume fraction.

\begin{figure}
\begin{center}
\mbox{\psfig{file=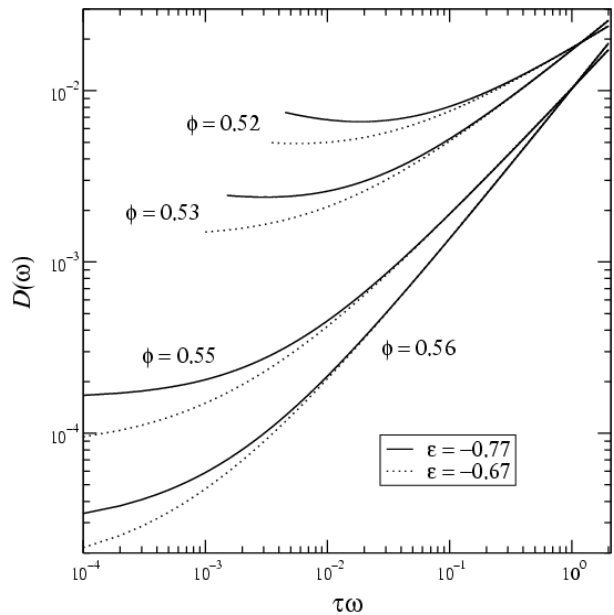,width=8cm}}
\caption{\label{DiffCH6B}\small Diffusion coefficient as a function of the frequency for
different values of the volume fraction. Solid lines were obtained by fitting the 
exponent $\epsilon =-0.77$, in accordance with Ref. \cite{bonn6}. Dotted lines 
were obtained by taking $\epsilon =-0.67$.}
\end{center}
\end{figure}

The expression of the diffusion coefficient at the frequencies we are considering
can now be derived by using Eq. (\ref{SE super}). In figure \textbf{2}, we
represent the diffusion coefficient given through Eq. (\ref{SE super}) as a
function of the reduced frequency. At higher concentrations (\( \phi =0.56,\, 0.55 \)),
a closer agreement with the experimental values of \( D \) given in \cite{bonn6}
has been obtained with \( \epsilon =-0.77 \) (solid lines). At lower concentrations
(\( \phi =0.53,\, 0.52 \)), a closer agreement with experiments has been obtained
with \( \epsilon =-0.67 \) (dotted lines). This dependence of the fitting on
the value of the exponent \( \epsilon  \) suggests that it is also a function
of the volume fraction.

\section{Conclusions}

In this paper, we have proposed a general formalism to analyze the dynamics
of systems which relax in long time scales of the order of the observation time
or even longer. The fact that the system always evolves through transient states
never reaching equilibration with the bath is the origin of a peculiar behaviour
different from that of systems whose propagating modes relax rapidly ending
up in a quiescent state. The dynamics is non-Markovian and the correlation functions
exhibit aging effects.

We have formulated a generalization of Onsager's theory to nonequilibrium aging
states to derive a Fokker-Planck equation which captures the main characteristics
of the dynamics such as non-stationarity through the two-time probability density,
non-Markovianity through the time-dependence of the transport coefficients and
lack of thermal equilibration between system and bath. \textbf{}This equation
is the main result of our analysis from which the behaviour of the correlation
functions follows. 

We have applied the theory to study the relaxation in supercooled colloidal
liquids for which violation of the Stokes-Einstein relation at low frequencies
has recently been found experimentally. The relation between diffusion coefficient
and viscosity we have obtained explains those experiments. The key ingredients
in this interpretation are the existence of a quasi-equilibrium state with a
temperature different from that of the bath which is proportional to the velocity
moments and the fact that the viscosity depends on the penetration length of
the diffusion modes resulting from the underlying activated process taking place
at sufficiently long times.

The formalism proposed thus offers an interesting theoretical framework for
the study of glassy behaviour in colloidal liquids.

\section{Acknowledgments}

We want to acknowledge Dr. D. Reguera for valious comments. I.S.H. acknowledges
UNAM-DGAPA for economic support. This work was partially supported by DGICYT
of the Spanish Government under Grant No. PB2002-01267.

\section{Acknowledgments}

\end{document}